# Mesoscopic MCT theory resolves Giant Non-Gaussian Parameter and Flory's conjecture

Yikun Ren[1*], Feixiang Xu[2], Ming Lin[1]

[1*]School of Materials Science and Engineering, Jiujiang University, Jiujiang, 332005, Jiangxi, China.
[2]BYD Company Limited, Shenzhen, 518000, Guangdong, China.
[*]Corresponding author(s). E-mail(s): 6260004@jju.edu.cn;
Contributing authors: fxxusz@gmail.com; linmingzny@163.com;

## Abstract:

Extending Prigogine's ideas to the interior of the system, we generalize mode-coupling theory from a microscopic to a mesoscopic formulation by incorporating the non-equilibrium eigen-phase. The resulting framework resolves two long-standing puzzles in glass transition physics with an error less than 0.01 against experiments: (i) the giant non-Gaussian parameter $\alpha_2 \sim 1 - 10$ which exceeds standard MCT predictions (only 0.1) by two orders of magnitude; (ii) the universal WLF constant $C_1 = 17ln10/(3\sqrt{42} - 19) \approx 16.7$, empirically observed for seven decades but never derived from first principles(e.g. Adam-Gibbs $C_1 = 8.5$, while other theories are off by more than a factor of two). These results establish mesoscopic MCT as a measurable foundation for non-equilibrium thermodynamics, unifying dynamic heterogeneity and thermodynamic universality in glass-forming systems.

## Introduction:

Experimentally, the non-Gaussian parameter $\alpha_2$ in glass-forming polymers, colloids and small-molecule liquids consistently falls in the range 1–10 [1–4]. Standard mode-coupling theory (MCT), which captures the local cage effect correctly [5], predicts $\alpha_2 \approx 0.1$ —an order-of-magnitude discrepancy that has persisted for decades. This large gap cannot be explained by refining the cage picture alone; it points to a mesoscopic entropy-driving mechanism that acts beyond the local scale and continuously reshapes dynamic heterogeneity.

In this work we identify such a mechanism by extending MCT to the mesoscopic level. Starting from the second law applied to a finite mesoscopic region, we show that the allowed microstates of such a region must form a contiguous subsequence of the full equilibrium microstate sequence—a truncation that defines the eigen-phase displacement r. The gradient of r acts as a conservative entropy-driving force that drives the system back toward its non-equilibrium eigen-phase. Incorporating this force into the Liouville equation via the Mori-Zwanzig formalism yields a generalized MCT equation whose restoring term is enhanced by a factor proportional to r. Solving this equation gives $\alpha_2 \in (1-10)$ without adjustable parameters, in quantitative agreement with experiments across multiple glass-forming systems.

Remarkably, the same mesoscopic framework also resolves a seemingly unrelated thermodynamic puzzle: the universal constant $C_1 \approx 16.7$ in the Williams–Landel–Ferry (WLF) equation. First proposed by Flory more than seventy years ago [6] and confirmed by countless experiments [7–9], $C_1$ has resisted any first-principles derivation. All equilibrium-based attempts—whether Simha-Boyer [10], Cohen-Grest [11] or Adam-Gibbs [12]—yield errors exceeding 100–300%, and the "free-volume" picture that once seemed to offer a qualitative explanation has long been abandoned as fundamentally ill-defined [13]. Here we show that the same eigen-phase displacement r, when evaluated at the glass transition, yields a critical vacancy fraction of 2.6% and directly gives $C_1 = 16.7$ with an error below 1%.

Two long-standing puzzles—one dynamic, one thermodynamic—are thus solved by a single mesoscopic theory. The eigen-phase displacement, far from being an abstract construct, emerges as a measurable quantity that unifies the description of non-equilibrium matter.

# I. Foundation of eigen-phase displacement theory

## 1.1 Theoretical Assumptions and Framework

For a nonequilibrium system, the thermodynamical structure of the eigen-phase are determined by the second law of thermodynamics through the following variational statement:

$$\delta(S(eigen\text{-}phase) + \int dS_e) = \delta(-k \sum_i P_i ln P_i + \int dS_e) = 0 \qquad (1)$$

In the equation (1), the $dS_e/dt$ is the external entropy flux, $S(eigen\text{-}phase)$ is the entropy of eigen-phase, P is the thermal probability of microstate i. The scheme of the eigen-phase "balance" could be shown in figure 1:

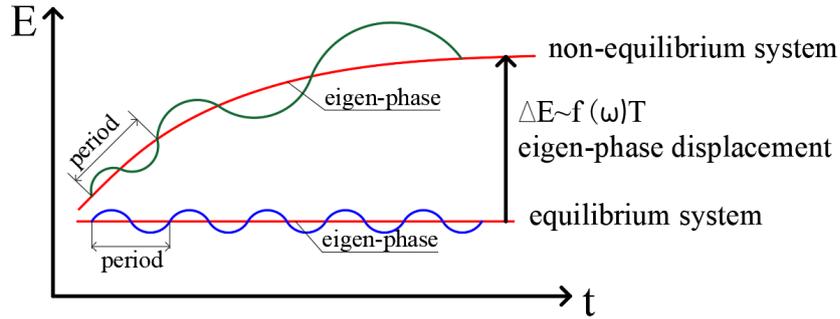

*Figure 1*.The fluctuation center manifests of equilibrium and non-equilibrium systems. The red line represents the fluctuation centers for equilibrium and non-equilibrim systems. The fluctuation center, fluctuation amplitude, and relaxation time of an equilibrium system remain constant. In contrast, the fluctuation center of the non-equilibrium system is displaced away from that of the equilibrium system, and both its fluctuation amplitude and relaxation time undergo changes. Due to fluctuations, an isolated non-equilibrium system will inevitably evolve toward equilibrium, but the process is typically much slower than predicted by dynamics alone. This is because the eigen-phase displacement, as illustrated in the figure, actively pushes particles in Mean Areas(defined in the following paragraph) that have deviated from the non-equilibrium eigen-phase back toward it, thereby retarding the relaxation process.

Building upon Prigogine's theory[14-16], we establish a new mesoscopic scale larger than the local scale of Prigogine. This larger mesoscopic scale is defined as follows: As we gradually zoom in from the macroscopic scale to the microscopic scale, there always exists a particular mesoscopic scale, denoted as the Mean Area (MA). When the observation scale is above the MA, the observer cannot detect dynamic heterogeneities but can observe the macroscopic statistical behavior of the system within a relatively short period of time. Conversely, when the observation scale is below the MA, the observer can detect dynamic heterogeneities but must observe over a longer period of time to grasp the macroscopic statistical behavior of the system. To provide a more rigorous mathematical and physical definition, based on the Local Assumption of Prigogine, we coarse-grain both the MA and the locale into N averaged particles (hereinafter referred to as "N-coarse-grained", where N is sufficiently large). The intersection of the sets of microstates of all locales is always contained within the set of microstates of the Mean Area (MA), and simultaneously, the set of microstates of the MA is always contained within the union of the sets of microstates of all locales. According to the definition of Mean Area, the mathematical definition of the Mean Area can be stated as: "The set of microstates of the coarse-grained MA is equal to the intersection of the sets of microstates of all coarse-grained locales" as shown in Equation (2).

$$\{microstates\ in\ CG\_MA\}= \bigcap_{locale \square MA} \{microstates\ in\ CG\_local\} \qquad (2)$$

By combining Equation (2) with the Local Assumption, it can be inferred that the N-coarse-grained Mean Area formally satisfies the partition function law. Without changing the degree of non-equilibrium, based on the second law of thermodynamics, the entropy of the CG_Mean Area could be regarded as "relatively equilibrium" if the locales are "small equilibrium systems". This is because, if the CG_Mean Area does not reach relative equilibrium without changing the degree of non-equilibrium, then CG-MA should be able to spontaneously relax towards any local. At a new moment, the new Spin generated in CG-MA will inevitably belong to a Spin that originally belonged only to a certain local, which violates equation (2). Therefore, a CG-MA that satisfies equation (2) will inevitably satisfy equation (1). The details of the framework could be seen in SI A.

## 1.2 The first part: MSS

From the second law of thermodynamics, it can be inferred that all states of the eigen-phases of any isolated system gradually increase in sequence, eventually reaching equilibrium. Therefore, decomposing the set of microstates of an equilibrium system into such a sequence, arranged in the order given by the arrow of time, can clearly reflect the changing of the set of non-equilibrium microstates. So, microstate sequence theory(MSS, briefly explained in SI B) is such a statistical theory built for calculation of eigen-phase [17]. Using the number-theoretical approaches and the second law of thermodynamics, MSS arrange all microstates in a continuous and monotonic sequence in order of energy or other non-equilibrium indices(e.g. in 3D Ising model, the index of any microstate is strictly written in the function of numbers of +1 spin-spin couplings and distance between neighbored +1). It is important to note that MSS is a mesoscopic theory for eigen-phase trajectory, not a complete representation of the system's microscopic phase trajectories. Using MSS theory and equation (1), here is the MSS CUTTING theorem(details see in SI B):

**All microstates of an eigen-phase of a Mean Area(MA) with certain energy always form a contiguous subsequence within the Microstate Sequence(MSS)**(The detailed deduction of MSS cutting theorem and exact expression of MSS in Ising model could be seen in SI A)**:**

$$P_i= \begin{cases} \dfrac{e^{-\beta E(i)}}{Z_{ne}} = \dfrac{e^{-\beta E(i)}}{\sum_{j=1}^{j=i_{max}} e^{-\beta E(j)}}, 1 \leq i \leq i_{max} \\ 0, i > i_{max} \end{cases} \qquad (3)$$

**{microstates' index in MSS cutting}={1, 2, 3,..., $i_{max}$}** (4)

The MSS Cutting Theorem gives explicit form to the non-equilibrium eigen-phase through a precise number-theoretic truncation procedure. **It breaks the ergodic hypothesis, yet restores ergodicity when the system reaches equilibrium.** As an extension of the second law of thermodynamics under the Law of Arrow of Time and Eigen-Phase, the theorem is important for understanding the thermodynamic behavior of non-equilibrium systems.

By combining Equations (3) and (4), the degree of non-equilibrium and eigen-phase displacement should be:

$$\omega = W/i_{max} \tag{5}$$

$$r = e^{\Delta E/kT} = \frac{\sum_{i=1}^{W/\omega} e^{-E(i)/kT}/(W/\omega)}{\sum_{i=1}^{W} e^{-E(i)/kT}/W} = \frac{\overline{P_{ne}}}{\overline{P_{eq}}} \tag{6}$$

$W$ is the number of microstates of a reference system (the equilibrium system is the first reference system), and $i_{max}$ represents both the maximum sequence number of microstates and the total number of microstates in the eigen phase. $\omega$ is the degree of non-equilibrium, which represents the degree of non-ergodicity of the whole system. E is the energy of microstate $i$. r represents the degree of non-ergodicity. $P_{ne}$ is the probability of a single particle choosing to be "non-equilibrium states". And $P_{eq}$ is the probability of a single particle choosing to be "equilibrium states". Therefore, $r$ also means the probability change for an equivalent mean single particle caused by the degree of non-equilibrium. Because it comes from the MSS cutting which origin in second law of thermodynamics, and we will see in the formulas of MCT that the r is the "generalized-force" **which drive the particles back to the non-equilibrium eigen-phase.** So, $r$ is defined as the eigen-phase displacement.

The entropy-driving force field originates from each Mean Area (MA) that is out of equilibrium, acting as a field source. This field drives the collective motion of particles both within and in the vicinity of the center of the MA, thereby sustaining the non-equilibrium state. The eigen-phase displacement function for a particle located at a position y, with $y-x=\epsilon$ relative to the center x of an MA, is given by(See in SI B)：

$$r(x+\epsilon) = \begin{cases} r_s(x)/2\epsilon_0, -\epsilon 0 \le \epsilon \le \epsilon 0 \\ 1, \epsilon > \epsilon 0 \end{cases} \tag{7}$$

$$lnr(q) = \sum_x lnr(x)(\theta(x+\epsilon)-\theta(x-\epsilon))e^{-iqx}\delta(x) \tag{8}$$

Equation (16) describes the degree of non-equilibrium of the MA with its central position at point of x. $\theta(x)$ is the Heaviside step function. 有趣的是，r 符合半群，而非李群，其详细推导见 SI B. Owing to this generalized non-local interaction, it can be inferred that the emergence of Cooperatively Rearranging Regions (CRR) of Adam-Gibbs theory is a necessary consequence, although a detailed discussion of this point lies beyond the scope of the present paper. The last important parameter is the eigen-phase displacement heterogeneity $a$(q), which means the potential field caused by the inertia force:

$$a(q) = \sum_i a(x_i)e^{-iqx_i} \approx iqSin(2\epsilon_0 q)a \approx iq^2 lnr \tag{9}$$

Note that substituting equation (6) into equation (1) yields a key equation governing the non-equilibrium phase transition.

$$E-ST-T\int_t^{t+\Delta t} Jdt \ge NkTlnr \tag{10}$$

It determines whether the non-equilibrium phase can still be maintained along its current relaxation path. While Prigogine's theory of open systems addressed how systems and their environment construct a "non-equilibrium balance", the concept of eigen-phase displacement explains why, within the system itself, each subsystem and even each particle tends to reside in a non-equilibrium phase beyond equilibrium, and why transitions in non-equilibrium relaxation pathways occur. The new theoretical framework aligns with Prigogine's thinking within the thermodynamic theoretical system, representing an inevitable consequence of the continuous driving by the second law of thermodynamics. Next, let's put the eigen-phase displacement into the **mescopic** Liouville equation to build the new MCT function.

## 1.3 the second part: MCT

**For mesoscopic particles**, considering that the eigen-phase displacement constitutes an obligatory expenditure of free energy, that is to say, the Hamiltonian must encompass the energy

associated with eigen-phase displacement. Consequently, the Liouville equation can be expressed as:

$$\frac{dC}{dt}=\{C,H(\omega)\}=\{C,H_0\}+\{C,\Delta E(x)\} \tag{11}$$

$$=iLC+\{C,k_BTlnr(x)\}=iLC-k_BT\sum_{i,j\neq i}\frac{\partial lnr(ij)}{\partial \vec{x_i}}\frac{\partial}{\partial \vec{p_j}}C=iLC-\Xi \tag{12}$$

For a given non-equilibrium system, the equation (12) could be simplified by the Mori-Zwanzig projection operator(details can be seen in SI C):

$$\frac{d^2F_{ne}(q,t)}{dt^2}+q^2k_BT/m(\frac{1}{S_f(q)}+aq^2)F_{ne}(q,t)+\frac{m}{Nk_BT}\int_0^t<R_qR_q(\tau)>\frac{dF_{ne}(q,t-\tau)}{dt}d\tau=0 \tag{13}$$

If $a=0$, which corresponds to zero heterogeneity in the degree of non-equilibrium and is physically equivalent to a zero degree of non-equilibrium, the equation (13) reduces to the MCT equations for equilibrium systems.

For nearly isolated systems (e.g., glass formers), the J≈0, so we get:

$$\begin{cases} E-ST\geq Nk_BTlnr \\ \frac{d^2F_{ne}(q,t)}{dt^2}+q^2k_BT/m(\frac{1}{S_f(q)}+aq^2)F_{ne}(q,t)+\frac{m}{Nk_BT}\int_0^t<R_qR_q(\tau)>\frac{dF_{ne}(q,t-\tau)}{dt}d\tau=0 \end{cases} \tag{14}$$

## II.Dynamical Theoretical verification: Non-Gaussain Parameters

Solving Equation (14) is largely analogous to solving a Newtonian mechanical equation; it necessitates introducing appropriate approximations to simplify the memory kernel $<R_{-q}R_q(\tau)>$. Different approaches, such as NMCT (Standard Mode-Coupling Theory)[5], GMCT (Generalized Mode-Coupling Theory)[18], EMCT (Extended Mode-Coupling Theory)[19], Elastic Theory of the Glass Transition[20], and NLE (Nonlinear Langevin Equation) theory[21], provide distinct pathways for this simplification.

In principle, methods like GMCT and EMCT offer more sophisticated treatments of the memory kernel. However, for addressing the core issue of eigen-phase, employing different simplification schemes does not alter its fundamental nature. Therefore, to most clearly highlight the critical role of the the a term, this paper employs the simplest NMCT approximation. The detailed solution can be found in SI C and SI D:

$$\alpha_2(peak-ne)=\lim_{q\to 0}\frac{2\left[\left(\frac{1}{2}-K\right)S_4+K\gamma S_0S_2+\frac{K\gamma^2S_0^3}{8}\right]}{\left[\left(\frac{1}{2}-K\right)S_2+\frac{K\gamma S_0^2}{2}\right]^2}-1 \tag{15}$$

In equation (15), $S_0=S_f(0)$, $S_2=\frac{1}{2}S_f''(0)$, $K=c_0\tau\sqrt{1-\frac{4}{\lambda}}$, $\gamma=\frac{4a}{\lambda-4}$, $\tau$ is the dynamic transition time.

If $a=0,r=1$, then the peak of NGP of MSS-MCT would return to the NGP of NMCT. This expression reveals a universal positive correlation between the non-Gaussian parameter $\alpha_2$ and the plateau value of the intermediate scattering function, Fne(0,τ). By using neutron scattering, SPT, DDM and computer simulations, this quantitative relationship aligns, both qualitatively and quantitatively, with experimental observations across various glass-forming systems such as small particles, colloids and polymers.[1-4]

The result is shown in figure 2 and Details could be seen in SI D. Crucially, our theory resolves the order-of-magnitude discrepancy in the non-Gaussian parameter: while standard NMCT yields 0.1, MSS-MCT correctly predicts $\alpha_2$ falls within the range of 1 to 10, in quantitative agreement with experiments across diverse glass-forming systems. Overall, the peak value of $\alpha_2$ hows a positive correlation with the intermediate function $K(peak) \propto \tau \propto 1/\langle \Delta x(t)^2 \rangle_{ne}$ and an inverse relationship with the mean-square displacement. This implies that when the cage is sufficiently tight, the relaxation time becomes long enough, consequently leading to a sufficiently high non-Gaussian parameter—a conclusion entirely consistent with the physical mechanism of the glass transition. 注意，无论是本征相位移（~0.04）还是结构因子（~0.001），它们都是非常小的数值，但两者耦合，就产生了如此巨大的非高斯参数，这正如非仿射运动理论[22]所揭示的那样。

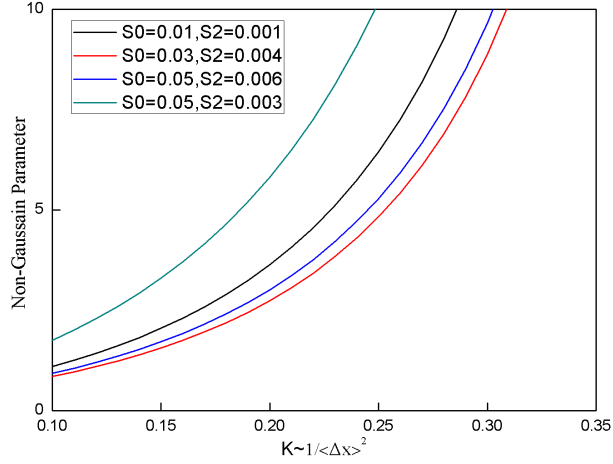

**Figure 2.** the relationship of NGP(peak) and K(peak) in MSS-MCT theory for systems with different structures. $K(peak) \propto \tau \propto 1/\langle \Delta x(t)^2 \rangle_{ne}$, So, NGP(peak) is inverse to mean square distance.

## III.Thermodynamical Theoretical verification: C1 in polymers

Using equation (14), we can get the $r$ parameter to build the entropy function of linear polymer. The detailed deduction is written in SI E. And the final expression of entropy is:

**Semi-Most-Probable Approximation**: Only the probability distribution parameter $r$ is retained to account for the role of MSS, while all microstates are treated as energetically equivalent.[23] $S$ is the entropy, $N$ is the number of all sites, $L$ is the number of segments in a single chain, $G$ is the number of chains, $r$ parameter is the eigen-phase displacement.

$$S = \frac{k}{r} ln \frac{N!}{(N-GLr)!} + kG(L-1)ln(\frac{5}{N}) \tag{16}$$

**Semi-Saddle-Point Approximation**: The full details of the energy ordering are preserved, but the partition function is treated within a semi-saddle-point approximation. This is equivalent to multiplying the probability P by a factor $\frac{Z_{ne}}{c_n^\rho}$(n is the total number of segment spins, $\rho$ represents the number of spins with a value of 6).

$$S \approx Q \frac{(GL-r)!r!}{(GLr)!(\frac{GL}{N}(N-GLr))!} \tag{17}$$

$Q$ is a non-divergent parameter whose expression is rather complex; its rigorous form is provided in SI E. These distinct approximation schemes demonstrate the **robustness** of the critical equation:

$$\frac{\partial S}{\partial r}\Big|_{N-GLr=0} = \infty \tag{18}$$

The two approximation schemes differ significantly in their mathematical and physical foundations—for instance, they yield substantially different expressions for the entropy function. However, the predicted critical point remains unaffected, thereby demonstrating the robustness of the conclusion. See Table 2.

| Robust | Assumption | Error bar of Probability | The key equation |
|---|---|---|---|
| Semi-Most-Probable Approximation | The eigen-phase displacements of all MA are assumed to be equal | $Z_{ne} = \sum_{i=1}^{W/\omega} e^{-E_i/kT} \approx k \ln W^{1/r}$ | $N-GLr=0$ |
| Semi-Saddle-Point Approximation | The partition function of each MA only held the collection at Saddle-point. | $Z_{ne} = \sum_{i=1}^{W/\omega} e^{-E_i/kT} \approx C_N^{N-pr} e^{-\overline{E_\rho}/kT}$ | $\frac{GL}{N}(N-GLr)=0$ |

Table 1, Robust certification of entropy "virtual-divergence" at transition point

Solving the critical point in equation (18), the result $C_1 = 17 \ln 10/(3\sqrt{42} - 19) \approx 16.7$(details see in SI F), let's compare it with experiments and classical theories.

From 1955, countless experiments have proved the correctness of $C_1$ by using DSC[24], DMA[25], PAT[26] and et al[27]. However, the non-fitting theoretical results is very few. So, we do not need to do more experiments to prove the $C_1$, but we should compare our theoretical result with the experiments and classical theories in the Table 3.

| Theory | Vacancy(%) | $C_1$ | Error vs Experiments | Hindered physics |
|---|---|---|---|---|
| Simha-Boyer (1962) | 11.6% | 3.7 | 455% | Vibration volume |
| Cohen-Grest (1979) | 7.6% | 5.7 | 300% | L-J potential |
| Adam-Gibbs (1965) | 5.3% | 8.2 | 207% | CRRS |
| MSS result (Our work) | 2.6% | 16.7 | 1% | Eigen-phase displacement |

Table 2. Theoretical and experimental values of $C_1$

From Table 3, our theory has the best accuracy to prove itself. The physical picture behind the 2.6% is the limit of displacement of eigen-phase. In another explanation, the transition happens on the critical point that the free energy of the non-equilibrium system were just enough to pay the "cost" of self-driving relaxation. For linear polymers, the Positron Annihilation Lifetime Spectroscopy (PALS) measures the glass transition temperature (Tg) by detecting the critical point at which the relative void fraction reaches approximately 2.5~2.6%. The Tg values obtained via PALS are consistent, within experimental error, with those measured by Differential Scanning Calorimetry (DSC) and other techniques. This agreement confirms that the glass transition temperature predicted by the MSS-MCT theory aligns with the experimentally observed Tg within the margin of error.

It is noteworthy that the calculation of $C_1$ within the MSS-MCT theory is confined to linear polymers and is valid only under conditions where the negative entropy flux from the environment is negligible. For certain polymer systems of higher complexity or under different experimental conditions, the **random packing** assumption may provide a more suitable description than the homogeneous volumetric expansion hypothesis for single-phase systems adopted in this work, leading to **different results** [28]. However, these outcomes do not violate the principles of mesoscopic dynamics or the second law of thermodynamics. Consequently, we conjecture that they do not contravene the MSS-MCT equations either, but merely necessitate more sophisticated treatment.

Lastly, if accuracy is everything, the Flory's conjecture is resolved in this paper. Now, the thermodynamics of Polymer glass transition has been explained by MSS-MCT theory to be a second-order non-equilibrium transition.

# V.Summary

We have shown that extending the second law of thermodynamics to finite mesoscopic regions leads to a fundamental quantity—the eigen-phase displacement r—which measures the deviation of a local non-equilibrium state from equilibrium. The gradient of lnr acts as a conservative entropy-driving force that counteracts fluctuations and sustains the eigen-phase. Incorporating this force into the Liouville equation via the Mori-Zwanzig formalism yields a generalized mode-coupling theory (MSS-MCT). The theory quantitatively resolves two long-standing puzzles in glass-forming materials:

1.The non-Gaussian parameter is predicted to lie in the range 1–10, correcting the order-of-magnitude underestimation of standard MCT and matching experiments across colloidal, small-molecule and polymeric systems.

2.The universal WLF constant of linear polymers is derived from first principles with less than 1% error, closing Flory's conjecture without invoking the ill-defined concept of free volume.

These two independent validations—one dynamic, one thermodynamic—establish the eigen-phase displacement as a measurable foundation for non-equilibrium thermodynamics. The framework naturally explains why equilibrium-based approaches fail and why the cage effect alone cannot account for strong dynamic heterogeneity.

The mathematical structure of the new force is a semi-group, encoding irreversibility in a way fundamentally different from Newtonian inertia. This opens the door to a unified description of non-equilibrium matter, from glassy relaxation to more complex phenomena such as dendritic segregation, the glass transition in ultrathin polymer films, and even the maintenance of living states. Future work will extend the theory to predict fragility and the full temperature dependence of relaxation times, and to explore its connection with energy-landscape approaches.

## Author Contribution:

**Yikun Ren**: **Conceived and developed the entire theoretical framework**, including the foundational concept of eigen-phase displacement; invented the Microstate Sequence (MSS) theory; derived all the SI; formulated the unified MSS-MCT equations; performed all formal analysis and analytical derivations; wrote the original manuscript.

**Feixiang Xu**: Provided resources, **assisted in clarifying the operational definition of the degree of non-equilibrium** within the proposed framework, assisted with literature review, and contributed to manuscript editing.

**Ming Lin**: Provided resources and contributed to manuscript editing.

## Acknowledge:

Prof. Rui Zhang named the MSS theory. Prof. Liangbin Li supported the whole work. Mr.Xiao Tu helped with the pictures. This work was supported by the National Natural Science Foundation of China (grant number 21973033), the Fundamental Research Funds for Central Universities (grant number 2018ZD13), Jiangxi Provincial Department of Education Project (grant number GJJ2401817), and the China Postdoctoral Science Foundation (grant number 2024M753321).

## Reference:

1   Flenner, E. & Szamel, G. Relaxation in a glassy binary mixture: Comparison of the mode-coupling theory to a Brownian dynamics simulation. Phys. Rev. E 72, 031508 (2005).

2   Mattsson, J., et al. Soft colloids make strong glasses. Nature 462, 83–86 (2009).

3   Vorselaars, B., Lyulin, A. V., Michels, M. A. J. & Karatasos, K. Non-Gaussian effects in the isotropic phase of a liquid crystal: Comparison of simulation with theory. Phys. Rev. E 75, 011504 (2007).

4   Habicht, J., et al. Non-Gaussian nature of glassy dynamics by cage to cage motion. J. Chem. Phys. 134, 184508 (2011).

5   Reichman, D. R. & Charbonneau, P. Mode-coupling theory. J. Stat. Mech. 2005, P05013 (2005).

6   Angell,C.A. Formation of glasses from liquids and biopolymers. Science 267(5206), 1924-1935.(1995)

7   Williams, M. L., Landel, R. F. & Ferry, J. D. The Temperature Dependence of Relaxation Mechanisms in Amorphous Polymers and Other Glass-forming Liquids. Journal of the American Chemical Society 77, 3701-3707 (1955).

8   Ferry, J. D. Viscoelastic Properties of Polymers. 3rd edn (Wiley, 1980).

9   Cohen, M. H. & Grest, G. S. Liquid-glass transition, a free-volume approach. Phys. Rev. B 20, 1077–1098 (1979).

10  Simha, R. , &   Boyer, R. F. On a general relation involving the glass temperature and coefficients of expansion of polymers. The Journal of Chemical Physics 37(5), 1003-1007 (1962).

11  Grest, G. S. & Cohen, M. H. Liquids, Glasses, and the Glass Transition: A Free-Volume Approach. Adv. Chem. Phys. 48, 455–525 (1981)

12  Adam, G. & Gibbs, J. H. On the Temperature Dependence of Cooperative Relaxation Properties in Glass-Forming Liquids. The Journal of Chemical Physics 43, 139-146 (1965).

13  Debenedetti, P. G., & Stillinger, F. H. Supercooled liquids and the glass transition. Nature 2001, 410(6825), 259–267.

14  Prigogine, I. Introduction to Thermodynamics of Irreversible Processes. 3rd edn (Wiley-Interscience, 1967).

15  De Groot, S. R. & Mazur, P. Non-Equilibrium Thermodynamics. (Dover Publications, 1984).

16  Maes, C. Non-Dissipative Effects in Nonequilibrium Systems. (Springer, 2018).

17  Ren, Y., Xu, F., Lin, M. & Hua, Q. Microstate Sequence Theory of Phase Transition: Theory Construction and Application on 3-Dimensional Ising Model. Fortschritte der Physik 73, 2300249 (2024).

18  Janssen, L. M. C., Mayer, P. & Reichman, D. R. Generalized mode-coupling theory of the glass transition. Phys. Rev. E 94, 054603 (2016).

19  Götze, W. & Sperl, L. Extended mode-coupling theory. Phys. Rev. E 66, 011405 (2002).

20  Zaccone, A. & Terentjev, E. M. Disorder-assisted melting and the glass transition in amorphous solids. Phys. Rev. B 88, 174203 (2013).

21  Schweizer, K. S. & Saltzman, E. J. Entropic barriers, activated hopping, and the glass transition in colloidal suspensions. J. Chem. Phys. 119, 1181–1196 (2003).

22  A. Zaccone & E. M. Terentjev, "Disorder-Assisted Melting and the Glass Transition in Amorphous Solids", Physical Review Letters 112, 105703 (2014).

23  Ren Y., Li Y., and Li L., A Theoretical Interpretation of Free Volume at Glass Transition,Chinese Journal of Polymer Science 35, 1415-1427(2017).

24  Wunderlich, B. Thermal analysis of polymeric materials. Thermochimica Acta 355, 43–57 (2000).

25  Jean, Y. C., et al. Characterization of free volume and density gradients in polymer films using positron annihilation spectroscopy. Macromolecules 40, 2522–2530 (2007).

26  Menard, K. P. Dynamic Mechanical Analysis: A Practical Introduction. 2nd ed. (CRC Press, 2008).

27  Mangialetto, J., Van den Brande, N., Van Mele, B., & Wübbenhorst, M. Real-Time Determination of the Glass Transition Temperature during Reversible Network Formation Based on Furan–Maleimide Diels–Alder Cycloadditions Using Dielectric Spectroscopy. Macromolecules, 56(12), 4727-4737 (2023).

28  Alghannam, E. M. A., Sabanayagam, C. R., Dittrich, P. S., & Isa, L. Molecular rotors to probe the local viscosity of a polymer glass. J. Chem. Phys. 156, 174901 (2022).